\documentclass[sigconf,article]{acmart}

\renewcommand\footnotetextcopyrightpermission[1]{} 
\usepackage{float}
\usepackage{booktabs} 
\usepackage{multirow}
\usepackage{makecell}
\usepackage{utfsym}
\usepackage{arydshln}
\usepackage{CJKutf8}
\usepackage{fancyvrb}
\usepackage{xcolor}
\usepackage{listings}
\usepackage{natbib}
\usepackage[skins,breakable]{tcolorbox}
\tcbset{
    colback=teal!5!white,
    sharp corners,
    enhanced,
    breakable=true,
    boxrule=1pt
}

\newtcolorbox{examplebox}[1][]{
    enhanced,
    breakable=true,
    colback=gray!10!white,
    sharp corners,
    boxrule=0.4pt,
    left=2mm,
    right=2mm,
    top=1mm,
    bottom=1mm,
    #1
}

\begin{document}

\title{KAP: MLLM-assisted OCR Text Enhancement for Hybrid Retrieval in Chinese Non-Narrative Documents}

\author{Hsin-Ling Hsu}
\affiliation{
    \institution{National Chengchi University}
    \country{Taipei, Taiwan}
}
\email{112306092@nccu.edu.tw}
\author{Ping-Sheng Lin}
\affiliation{
    \institution{National Chengchi University}
    \country{Taipei, Taiwan}
}
\email{111307050@nccu.edu.tw}
\author{Jing-Di Lin}
\affiliation{
    \institution{National Chengchi University}
    \country{Taipei, Taiwan}
}
\email{111301029@nccu.edu.tw}
\author{Jengnan Tzeng$^{*}$}
\affiliation{
    \institution{National Chengchi University}
    \country{Taipei, Taiwan}
}
\email{glophy@g.nccu.edu.tw}

\begin{abstract}
Hybrid Retrieval systems, combining Sparse and Dense Retrieval methods, struggle with Traditional Chinese non-narrative documents due to their complex formatting, rich vocabulary, and the insufficient understanding of Chinese synonyms by common embedding models. Previous approaches inadequately address the dual needs of these systems, focusing mainly on general text quality improvement rather than optimizing for retrieval. We propose Knowledge-Aware Preprocessing (KAP), a novel framework that transforms noisy OCR outputs into retrieval-optimized text. KAP adopts a two-stage approach: it first extracts text using OCR, then employs Multimodal Large Language Models to refine the output by integrating visual information from the original documents. This design reduces OCR noise, reconstructs structural elements, and formats the text to satisfy the distinct requirements of sparse and dense retrieval. Empirical results demonstrate that KAP consistently and significantly outperforms conventional preprocessing approaches. Our code is available at \href{https://github.com/JustinHsu1019/KAP}{https://github.com/JustinHsu1019/KAP}.

\end{abstract}


\maketitle
\pagestyle{plain} 

\section{Introduction}
Retrieval-Augmented Generation (RAG) \cite{NEURIPS2020_6b493230} has emerged as a key technology for knowledge-driven applications, enabling systems to retrieve relevant information from large-scale knowledge bases to enhance tasks such as question answering and decision support. However, when dealing with non-narrative documents (e.g., financial statements, contractual clauses, and tables embedded in PDFs), traditional RAG-based retrieval methods face significant challenges due to poor OCR quality, loss of document structure, and suboptimal text chunking.

A major issue arises from OCR processing errors, particularly in Traditional Chinese financial documents, where character misrecognition, formatting loss, and disrupted table structures degrade retrieval accuracy. Common OCR techniques struggle to preserve tabular relationships, causing critical numerical and textual data to be misaligned or fragmented. Additionally, existing chunking methods are not optimized for non-narrative text, often splitting semantically related content incorrectly, further reducing retrieval performance.

One of the most widely adopted retrieval methods today is Hybrid Retrieval \cite{Ma2020HybridFR,azureai2023}, which combines Sparse Retrieval (e.g., BM25) and Dense Retrieval to leverage both exact keyword matching and semantic search. Sparse Retrieval methods like BM25 \cite{bm251994} rank documents based on term frequency-inverse document frequency (TF-IDF) and document length normalization, making them efficient for queries where keyword overlap strongly indicates relevance. Meanwhile, Dense Retrieval \cite{karpukhin-etal-2020-dense} encodes queries and documents into a shared embedding space for semantic similarity search, capturing contextual relationships beyond exact matches. Nevertheless, the effectiveness of both methods depends heavily on the quality of the retrieved text. Dense Retrieval models perform poorly on noisy OCR output and struggle to capture meaning from fragmented text. Meanwhile, Sparse Retrieval (BM25) relies on exact term matching, making it sensitive to synonym variations and phrasing differences, particularly in Traditional Chinese, where word order plays a crucial role in retrieval performance, and the language’s vast vocabulary with abundant synonyms makes semantic matching even more challenging for retrieval systems.

While existing post-OCR processing methods attempt to address character recognition errors and text correction \cite{10.1145/3453476}, they are fundamentally designed for general text quality improvement rather than optimizing for retrieval systems. These methods typically focus on linguistic correctness and readability, overlooking the specific requirements of retrieval mechanisms where term frequency, keyword distribution, and semantic coherence are critical for performance. Current correction approaches improve textual accuracy but fail to consider how the processed text will ultimately be consumed by retrieval systems. This disconnect creates a significant gap—the ultimate purpose of knowledge preprocessing is to facilitate retrieval, yet existing methods do not adapt their strategies to the specific requirements of different retrieval approaches.

To address these challenges, we propose Knowledge-Aware Preprocessing (KAP), a two-stage preprocessing framework that optimizes textual representations specifically for Hybrid Retrieval in non-narrative documents. Rather than applying generic text improvement techniques, KAP recognizes that knowledge preprocessing must be tailored to the downstream retrieval task. Our framework enhances input data quality by leveraging Multimodal Large Language Models (MLLMs) with LLM-driven post-OCR processing that is explicitly designed to improve retrieval performance. KAP refines extracted text, corrects OCR errors, restores lost structures, and strategically restructures content to improve compatibility with both Sparse and Dense Retrieval components of Hybrid systems.

The main contributions of this study are as follows:
\begin{itemize}
    \item \textbf{Proposing the KAP framework}: A two-stage MLLM-based preprocessing approach that corrects OCR errors, reconstructs table structures, optimizes text representation in non-narrative documents, and enhances compatibility with both Sparse and Dense Retrieval.
    \item \textbf{Empirical validation of KAP's effectiveness}: Tests on the E.SUN Bank dataset demonstrate that our framework significantly improves retrieval accuracy and structure preservation in Traditional Chinese financial documents.
    \item \textbf{Designing specialized validation methodology}: Creating an LLM-driven approach for augmenting validation datasets specifically optimized for Hybrid Retrieval evaluation, generating diverse query variations that comprehensively test both sparse and dense retrieval components.
\end{itemize}

\section{Related Work}

Optical Character Recognition (OCR) technology has been widely applied to text extraction from scanned documents and PDF files, but still faces significant challenges when processing non-narrative documents such as financial statements, contracts, and tables. These challenges include character recognition errors, format loss, and structural misalignment, which hinder accurate retrieval of content. Existing post-OCR processing methods generally fall into three categories: manual correction (accurate but labor-intensive), isolated-word methods (dictionary-based approaches that struggle with context), and context-dependent methods (from feature-based techniques to sequence-to-sequence models) \cite{10.1145/3453476}. Recent advancements have introduced sequence-to-sequence \cite{seq2seq} and Transformer-based models \cite{Li_Lv_Chen_Cui_Lu_Florencio_Zhang_Li_Wei_2023} that improve text correction, while Large Language Models (LLMs) further enhance OCR correction by incorporating textual and structural understanding \cite{soper-etal-2021-bart}, but they are not optimized for the preferences of retrieval systems where term frequency and semantic coherence are critical.

In Retrieval-Augmented Generation (RAG) and Hybrid Retrieval systems, text quality significantly impacts retrieval outcomes. Sparse Retrieval methods (e.g., BM25) are sensitive to spelling errors and formatting disruptions, while Dense Retrieval methods perform poorly with fragmented text lacking context—particularly evident when handling table data. Previous studies have proposed improvements such as Contextual Retrieval \cite{anthropic_contextual_retrieval} and Structure-Aware Transformers for table verification \cite{zhang_etal_2020_table}, but these primarily focus on semantic modeling rather than text preprocessing optimization.

The KAP framework proposed in this study builds on MLLM-based post-OCR techniques and is specifically optimized for Hybrid Retrieval requirements. Unlike traditional OCR correction methods, KAP incorporates a prompt-based approach that ensures generated text formats are suitable for retrieval systems, leverages LLMs for contextual understanding and structural restoration of tables, and precisely controls text structure and keyword distribution to enhance both Sparse and Dense Retrieval performance.

\section{Methods}

This section presents the proposed KAP framework, a two-stage preprocessing pipeline designed to enhance the effectiveness of Hybrid Retrieval in non-narrative documents. The framework addresses challenges associated with processing non-narrative documents, particularly financial statements and contractual clauses, which often contain complex tabular structures and require precise semantic preservation. 

\begin{figure*}[htbp]
    \centering
    \includegraphics[width=1\textwidth]{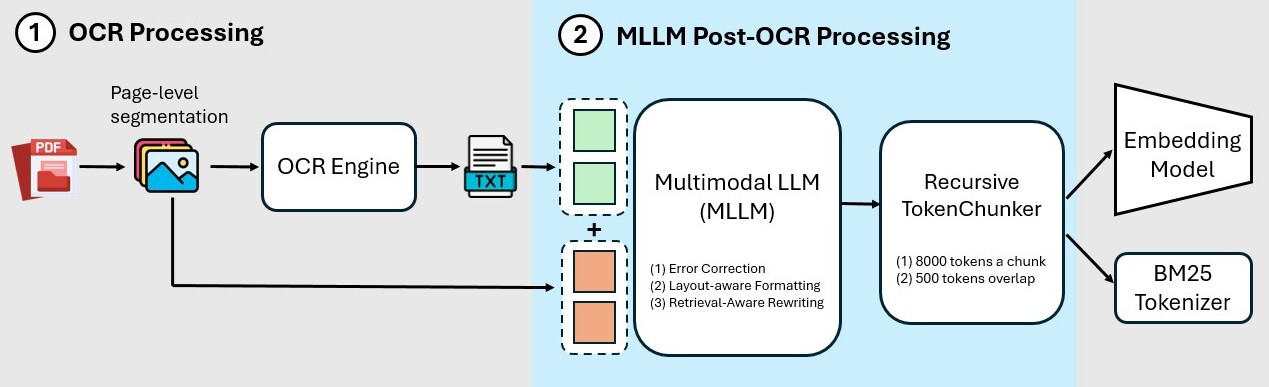}
    \caption{Overall architecture of the proposed KAP framework.}
    \label{fig:model}
\end{figure*}

KAP consists of two main stages: 
(1) OCR Processing, which extracts text from PDFs, and 
(2) MLLM Post-OCR Processing, which refines the extracted text for better compatibility with Sparse and Dense Retrieval. 
By leveraging Multimodal Large Language Models (MLLMs) and prompt engineering, KAP corrects OCR errors, restores table structures, and optimizes text format to enhance retrieval accuracy. Figure~\ref{fig:model} illustrates the overall architecture of KAP.

\subsection{OCR Processing}

The first stage of KAP involves extracting textual content from PDFs using Optical Character Recognition (OCR).

While OCR facilitates text extraction, it also introduces common errors, particularly when handling scanned financial documents. Character misrecognition often occurs, with errors such as "0" being recognized as "O" or "1" as "l." Additionally, the extraction process results in the loss of table structures, causing the extracted text to lack the original tabular layout and disrupting data relationships.

To mitigate these issues, we introduce an LLM-driven post-processing stage that refines OCR output and ensures compatibility with retrieval systems.

\subsection{MLLM Post-OCR Processing}

The second stage of KAP employs Multimodal Large Language Models (MLLMs) to enhance OCR-extracted text quality. Through prompt engineering with a single prompt, KAP effectively corrects OCR errors, reconstructs table structures, and optimizes text formatting to improve retrieval accuracy.

As illustrated in Figure~\ref{fig:prompt}, the prompt template takes as input the OCR-extracted text along with the corresponding original image, guiding the MLLM to perform targeted improvements. The prompt was originally written in Traditional Chinese; its English translation is provided at the beginning of each subsection in this section.

This process consists of three key components: error correction, layout-aware format reconstruction, and retrieval-aware rewriting. This streamlined approach simplifies the entire workflow described in this section while maintaining high performance across all processing tasks.

\subsubsection{Error Correction}

As shown in Figure \ref{fig:prompt}, the first component (red box, lines 5-6) of our prompt template:

\begin{examplebox}[breakable]
1. OCR Error Correction \\
- Correct errors in OCR conversion (e.g., typos, omissions, inverted sentence order, etc.), making the text more fluent and grammatically correct.
\end{examplebox}

OCR outputs often contain spelling mistakes, incorrect numerical values, and misplaced punctuation. To address this, we apply an LLM-based correction mechanism that fixes common OCR recognition errors in Traditional Chinese and ensures grammatical correctness and readability while preserving semantic meaning.

\subsubsection{Layout-Aware Format Reconstruction}

As shown in Figure \ref{fig:prompt}, the component highlighted in the blue box (lines 38) of our prompt template:

\begin{examplebox}[breakable]
You can refer to the attached image to help you understand how this text is presented in the original PDF file (e.g., tables, narrative sentences), and what each text and number represents and where they appear in the original text.
\end{examplebox}

Financial documents often rely on structured layouts, such as tables, section headers, and alignment, which standard OCR processing does not fully capture. To enhance text representation, KAP leverages MLLM's multimodal capabilities within LLM-driven post-OCR processing. Specifically, MLLM's vision capability is used to interpret the original PDF layout and refine extracted text accordingly, ensuring tables and formatting are preserved.

\subsubsection{Retrieval-Aware Rewriting}

\begin{figure*}[htbp]
    \centering
    \includegraphics[width=0.9\textwidth]{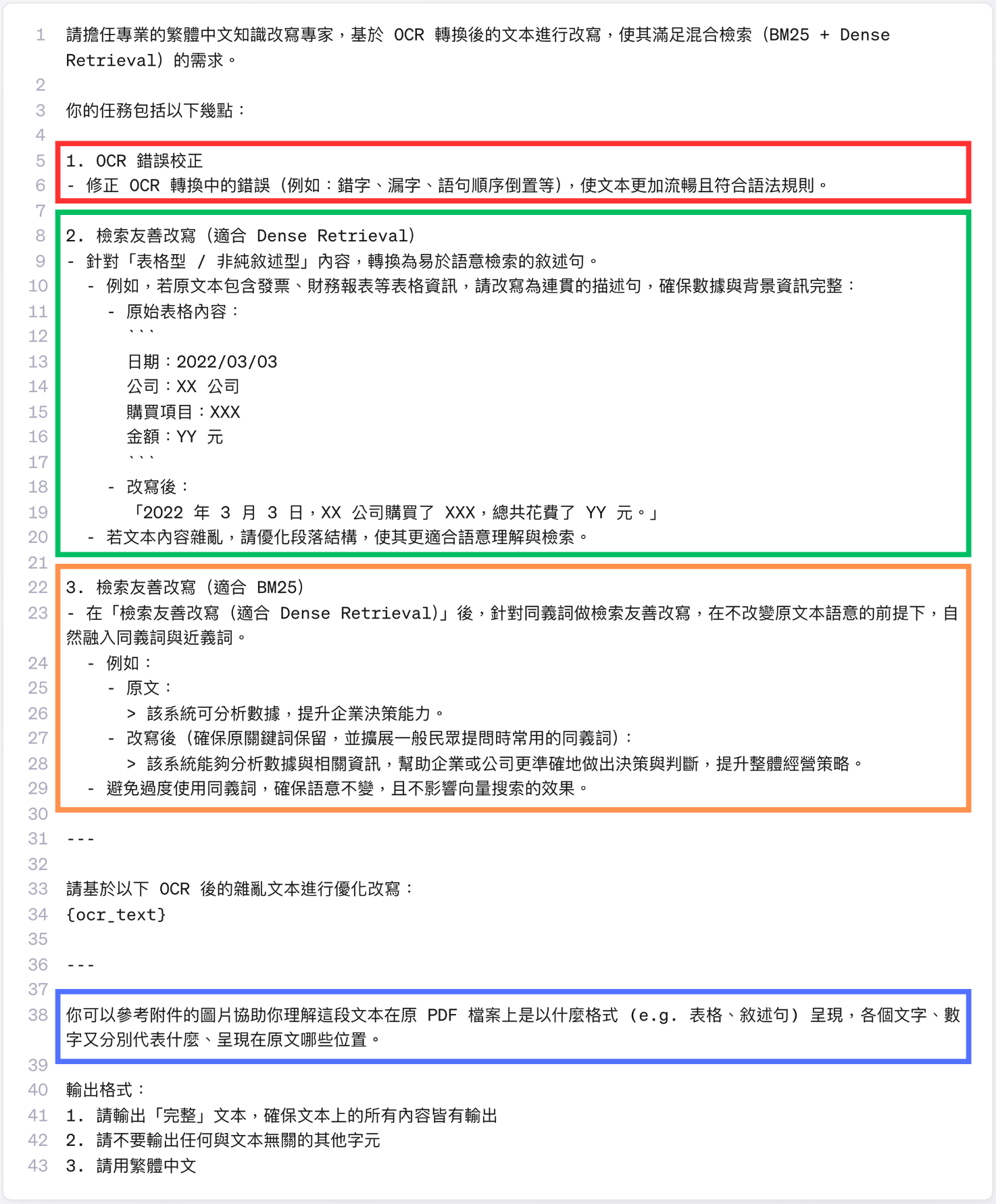}
    \caption{The prompt template used for post-OCR processing with MLLM, originally written in Traditional Chinese. Its English translation is provided in the subsections of Section \textit{3.2 MLLM Post-OCR Processing}.}
    \label{fig:prompt}
\end{figure*}

KAP optimizes extracted text to improve compatibility with Hybrid Retrieval methods, ensuring better alignment with both Sparse and Dense retrieval strategies:

\begin{enumerate}
    \item Dense Retrieval Optimization: As shown in Figure \ref{fig:prompt}, the component highlighted in the green box (lines 8-20) of our prompt template:

    \begin{examplebox}[breakable]
    2. Retrieval-Friendly Rewriting (Suitable for Dense Retrieval) \\
    - For "tabular/non-pure narrative" content, convert into semantic retrieval-friendly narrative sentences. \\
    \hspace*{1em}- For example, if the original text contains invoice, financial statement, or other tabular information, please rewrite into coherent descriptive sentences, ensuring the data and background information are complete: \\
    \hspace*{1em}\hspace*{1em}- Original tabular content: \\
    \hspace*{1em}\hspace*{1em}\hspace*{1em}``` \\
    \hspace*{1em}\hspace*{1em}\hspace*{1em}Date: 2022/03/03 \\
    \hspace*{1em}\hspace*{1em}\hspace*{1em}Company: XX Company \\
    \hspace*{1em}\hspace*{1em}\hspace*{1em}Purchase item: XXX \\
    \hspace*{1em}\hspace*{1em}\hspace*{1em}Amount: YY dollars \\
    \hspace*{1em}\hspace*{1em}\hspace*{1em}``` \\
    \hspace*{1em}\hspace*{1em}- After rewriting: \\
    \hspace*{1em}\hspace*{1em}\hspace*{1em}"On March 3, 2022, XX Company purchased XXX, spending a total of YY dollars." \\
    \hspace*{1em}- If the text content is messy, please optimize the paragraph structure to make it more suitable for semantic understanding and retrieval.
    \end{examplebox}
    
    Reformulates structured tabular content into natural language descriptions to improve semantic embedding quality, ensuring better performance in embedding models:
    \begin{itemize}
        \item Original table:
        \begin{table}[h]
            \centering
            \begin{tabular}{|c|c|c|}
            \hline
            \begin{CJK}{UTF8}{bkai}日期\end{CJK} (Date) & \begin{CJK}{UTF8}{bkai}公司\end{CJK} (Company) & \begin{CJK}{UTF8}{bkai}金額\end{CJK} (Amount) \\
            \hline
            2025/03/03 & XX & \$10,000 \\
            \hline
            \end{tabular}
        \end{table}
        \item Reformulated as:
        \begin{tcolorbox}[breakable]
        \begin{CJK}{UTF8}{bkai}  
        2025年3月3日，XX 公司記錄了一筆 \$10,000 的交易。  
        \end{CJK} \\ (On March 3, 2025, XX Corp recorded a transaction of \$10,000.)  
        \end{tcolorbox}
    \end{itemize}
    \item Sparse Retrieval (BM25) Optimization: As shown in Figure \ref{fig:prompt}, the component highlighted in the orange box (lines 22-29) of our prompt template:

    \begin{examplebox}[breakable]
    3. Retrieval-Friendly Rewriting (Suitable for BM25) \\
    - After "Retrieval-Friendly Rewriting (Suitable for Dense Retrieval)," rewrite synonyms in a retrieval-friendly way, naturally incorporating synonyms and near-synonyms while maintaining the original text's meaning. \\
    \hspace*{1em}- For example: \\
    \hspace*{1em}\hspace*{1em}- Original text: \\
    \hspace*{1em}\hspace*{1em}\hspace*{1em}> The system can analyze data, enhancing the enterprise's decision-making ability. \\
    \hspace*{1em}\hspace*{1em}- After rewriting (ensuring the original keywords are preserved and expanding synonyms commonly used by the general public when asking questions): \\
    \hspace*{1em}\hspace*{1em}\hspace*{1em}> The system is able to analyze data and related information, helping enterprises or companies make more accurate decisions and judgments, improving overall business strategy. \\
    \hspace*{1em}- Avoid overusing synonyms, ensure the meaning remains unchanged, and do not affect the effect of vector search.
    \end{examplebox}
    
    Expands key terms with synonyms to improve keyword matching while ensuring important keywords remain intact for query relevance.

    \begin{itemize}
        \item Original:
        \begin{tcolorbox}[breakable]
        \begin{CJK}{UTF8}{bkai}  
        系統可以分析數據，以提升商業決策能力。  
        \end{CJK} \\ (The system can analyze data to enhance business decision-making.)  
        \end{tcolorbox}
        \item Rewritten (with synonym expansion while retaining key terms):
        \begin{tcolorbox}[breakable]
        \begin{CJK}{UTF8}{bkai}  
        系統能夠分析數據與相關資訊，幫助企業或公司做出更精確的決策，並提升整體策略。  
        \end{CJK} \\ (The system is capable of analyzing data and relevant information, helping businesses or companies make more accurate decisions and improve overall strategy.)  
        \end{tcolorbox}
    \end{itemize}
    \textbf{Expansion:} \begin{CJK}{UTF8}{bkai}Added synonyms for "數據 (data)" (資訊 for information) and elaborated on "決策 (decision)" (策略 for strategy).\end{CJK}
\end{enumerate}

\subsection{Chunking Strategy}

Traditional chunking methods such as Recursive Chunking \cite{weaviate_recursive_chunking} are not optimized for non-narrative documents, often splitting critical tabular content across multiple chunks. To mitigate this issue, KAP adopts a two-step chunking strategy:
\begin{enumerate}
    \item Page-level segmentation: Documents are first segmented based on page boundaries to preserve contextual relationships.
    \item Recursive Chunking: After MLLM Post-OCR processing, text is further segmented using Recursive Chunking with optimized chunk size (8,000 tokens with 500-token overlap) to prevent excessive fragmentation.
\end{enumerate}

\section{Experiments}

\subsection{Experimental Setup}

\subsubsection{Datasets and Preprocessing}

This study uses a non-public dataset from E.SUN Bank, provided through the "AI CUP 2024 E.SUN Artificial Intelligence Open Competition." The dataset contains three types of documents: FAQ, insurance terms, and financial reports. This study primarily focuses on financial reports since they contain a large number of tables, aligning with our research goal of handling non-narrative text. In contrast, the FAQ category is stored in JSON format with simple question-answer pairs, and the insurance terms mainly consist of plain text contracts, which differ from the complex formatted texts that this study targets. Therefore, these categories are not included in the main analysis.

Table \ref{tab:dataset} provides an overview of the document formats and contents in the E.SUN Bank dataset.

\begin{table}[h]
    \centering
    \caption{Overview of E.SUN Bank Dataset Categories}
    \label{tab:dataset}
    \resizebox{\columnwidth}{!}{%
    \begin{tabular}{lllc}
        \toprule
        Category & Description & Docs & File Type \\
        \midrule
        FAQ & Frequently asked questions & 617 & JSON \\
        Insurance & Policy terms sold by E.SUN Bank & 643 & PDF \\
        Finance & Financial reports of listed companies & 1035 & PDF \\
        \bottomrule
    \end{tabular}
    }
\end{table}

The validation set, provided by E.SUN, contains multiple questions, each associated with a document that provides the answer. Each entry includes a question ID, query content, data source, and category. For each category (FAQ, Insurance, Finance), the validation set includes 50 questions. Table \ref{tab:query_format} summarizes the format of the validation set fields.

\begin{table}[h]
    \centering
    \caption{Validation Set Question Format}
    \label{tab:query_format}
    \begin{tabular}{ll}
        \toprule
        Field & Description \\
        \midrule
        qid & Question ID \\
        query & Question content \\
        source & Candidate Document ID (pid) \\
        category & Data type (FAQ / Insurance / Finance) \\
        \bottomrule
    \end{tabular}
\end{table}

To enhance the diversity of the validation set and rigorously evaluate the retrieval model's robustness under various conditions, we expand the original set of 50 Q\&A pairs to 500 questions through systematic data augmentation. Specifically, for each original question, we generate 9 distinct reformulations using the following augmentation strategies:

\begin{enumerate}
    \item Replacing all keywords in the question with their synonyms (e.g., ``modify'' $\rightarrow$ ``change'').
    \item Substituting half of the keywords with synonyms while keeping the other half unchanged.
    \item Extracting the core keywords from the question and presenting them as a space-separated list.
    \item Extracting the core keywords and replacing all of them with synonyms, formatted as a space-separated list.
    \item Extracting the core keywords and replacing only half of them with synonyms, maintaining the rest as originally phrased, in a space-separated format.
    \item Transforming sentence structure by altering the word order, such as swapping the subject and verb while preserving the original meaning.
    \item Condensing the question into its most concise form while retaining its essential meaning.
    \item Applying structural variations to the condensed version, such as reordering the subject and verb.
    \item Reformulating the question into a more informal and conversational style.
\end{enumerate}

To generate these augmented question variants, we leverage a large language model (Claude-3.7-Sonnet \cite{anthropic_claude3_sonnet}) by providing structured prompt that systematically apply the above transformation techniques.

\subsubsection{Model Selection}

For our OCR processing, we utilize Tesseract OCR \cite{4376991}, the most commonly used open-source engine with robust support for Traditional Chinese. This selection ensures reliable text extraction from the PDF documents in our dataset.

In the MLLM post-OCR processing stage, we employ Claude-3.7-Sonnet \cite{anthropic_claude3_sonnet} as our multimodal language model. This model was selected for its strong multimodal text-image understanding capabilities, which are essential for correcting OCR errors, reconstructing table structures, and optimizing text formatting. The same model was also leveraged for generating our augmented validation questions, ensuring consistency throughout our experimental pipeline.

\subsection{Metrics}

To comprehensively evaluate retrieval performance, this study adopts the following standard metrics. Mean Reciprocal Rank (MRR) measures the rank of the first correct answer in the retrieval results, where a higher value indicates that correct answers tend to appear higher in the model's ranked results. Average Precision@1 measures whether the top 1 retrieved result is the correct answer, with a higher value indicating that the model can accurately locate key information.

\subsection{Evaluation Methodology}
To evaluate our KAP preprocessing framework, we measure improvements across three retrieval strategies:
\begin{enumerate}
    \item \textbf{Sparse Retrieval (BM25)}: 
    Ranks documents based on keyword matching using term frequency-inverse document frequency and document length normalization. In our implementation, we use the Jieba \cite{jieba} tokenizer with a Traditional Chinese lexicon.
    \item \textbf{Dense Retrieval}:  
    Encodes queries and documents into a shared embedding space—using the text-embedding-3-large \cite{openai2024textembedding3large} model—allowing for semantic similarity search rather than exact keyword matching.
    \item \textbf{Hybrid Retrieval}: 
    Combines both sparse (BM25) and dense retrieval methods to leverage the strengths of keyword-based and semantic-based retrieval, capturing both exact matches and deeper semantic relevance.
\end{enumerate}

\subsubsection{Baseline System}

We use \textbf{Tesseract OCR} as the baseline system, which directly applies OCR-extracted text for retrieval without additional preprocessing.

\subsubsection{Ablation Study}

To evaluate the impact of different components in the preprocessing framework, we perform ablation experiments on our KAP system. The test configurations include:
\begin{enumerate}
    \item \textbf{KAP w/o Vision}: Removing MLLM’s ability to parse images (tables, layout), using only OCR text results to test whether MLLM improves format reconstruction.
    \item \textbf{KAP w/o OCR Text}: Removing OCR-converted text and relying solely on MLLM to extract content from PDF images to test the influence of text input.
    \item \textbf{KAP w/o Rewrite}: Removing the text rewriting module in the KAP framework, which aligns text formats for Dense and Sparse retrieval, to test its impact on retrieval accuracy.
    \item \textbf{Full KAP Framework (Ours)}: Providing complete OCR text, MLLM parsing, and text rewriting to ensure optimal output for Hybrid Retrieval.
\end{enumerate}

\subsection{Results}

Each experiment was independently repeated three times. We report the mean and standard deviation across the three runs to ensure consistency and robustness. Tables~\ref{tab:bm25_results}, \ref{tab:dense_results}, and \ref{tab:hybrid_results} summarize the retrieval performance under Sparse, Dense, and Hybrid settings, respectively. As shown, our proposed KAP framework consistently outperforms the baseline across all retrieval methods.

\begin{table}[h]
    \centering
    \caption{Performance of Sparse Retrieval (Mean $\pm$ SD)}
    \label{tab:bm25_results}
    \begin{tabular}{lcc}
        \toprule
        Methods & MRR (\%) & Precision@1 (\%) \\
        \midrule
        Tesseract OCR (Baseline) & 53.16$\pm$0.83 & 41.51$\pm$1.67 \\
        \midrule
        KAP w/o Vision & 54.84$\pm$1.24 & 43.66$\pm$1.45 \\
        KAP w/o OCR Text & 62.32$\pm$0.43 & 49.39$\pm$0.73 \\
        KAP w/o Rewrite & 59.60$\pm$1.03 & 45.45$\pm$1.56 \\
        \midrule
        KAP (Ours) & \textbf{63.64$\pm$0.09} & \textbf{51.16$\pm$0.21} \\
        \bottomrule
    \end{tabular}
\end{table}

\begin{table}[h]
    \centering
    \caption{Performance of Dense Retrieval (Mean $\pm$ SD)}
    \label{tab:dense_results}
    \begin{tabular}{lcc}
        \toprule
        Methods & MRR (\%) & Precision@1 (\%) \\
        \midrule
        Tesseract OCR (Baseline) & 48.41$\pm$0.60 & 32.10$\pm$0.74 \\
        \midrule
        KAP w/o Vision & 56.62$\pm$0.39 & 42.98$\pm$0.65 \\
        KAP w/o OCR Text & 54.00$\pm$0.63 & 44.41$\pm$0.47 \\
        KAP w/o Rewrite & 58.46$\pm$1.32 & 46.11$\pm$1.65 \\
        \midrule
        KAP (Ours) & \textbf{65.16$\pm$1.51} & \textbf{53.65$\pm$2.24} \\
        \bottomrule
    \end{tabular}
\end{table}

\begin{table}[h]
    \centering
    \caption{Performance of Hybrid Retrieval (Mean $\pm$ SD)}
    \label{tab:hybrid_results}
    \begin{tabular}{lcc}
        \toprule
        Methods & MRR (\%) & Precision@1 (\%) \\
        \midrule
        Tesseract OCR (Baseline) & 53.23$\pm$0.57 & 38.98$\pm$0.88 \\
        \midrule
        KAP w/o Vision & 58.52$\pm$0.51 & 47.33$\pm$0.81 \\
        KAP w/o OCR Text & 65.06$\pm$0.15 & 56.39$\pm$0.11 \\
        KAP w/o Rewrite & 66.02$\pm$1.71 & 55.48$\pm$2.13 \\
        \midrule
        KAP (Ours) & \textbf{69.46$\pm$0.61} & \textbf{59.73$\pm$1.10} \\
        \bottomrule
    \end{tabular}
\end{table}

\subsubsection{Comparison with Baseline}

For Sparse Retrieval (BM25), KAP improves MRR from 53.16\%$\pm$0.83 to 63.64\%$\pm$0.09 and Precision@1 from 41.51\%$\pm$1.67 to 51.16\%$\pm$0.21. This improvement is attributed to better text normalization and structural refinements, which enhance keyword matching effectiveness. In contrast, the baseline struggles with OCR errors and unstructured text, resulting in lower accuracy.

In Dense Retrieval, KAP outperforms the baseline by increasing MRR from 48.41\%$\pm$0.60 to 65.16\%$\pm$1.51 and Precision@1 from 32.10\%$\pm$0.74 to 53.65\%$\pm$2.24. The significant performance gain highlights the role of text refinement in improving vector-based retrieval, as cleaner input text leads to better semantic embeddings. The baseline's noisy OCR output negatively impacts dense retrieval performance.

Hybrid Retrieval benefits the most from KAP, achieving the highest accuracy with an MRR improvement from 53.23\%$\pm$0.57 to 69.46\%$\pm$0.61 and Precision@1 from 38.98\%$\pm$0.88 to 59.73\%$\pm$1.10. This suggests that KAP effectively enhances both keyword-based and semantic retrieval components, leading to a more robust retrieval pipeline. The baseline's poor text quality affects both BM25 and vector search components.

\subsubsection{Ablation Study}

We conduct an ablation study to evaluate the impact of different components in the KAP framework. Across all retrieval strategies, removing any component results in performance degradation, with the most significant drops consistently occurring when vision processing is removed.

For Sparse Retrieval, the largest decrease occurs with KAP w/o Vision, reducing Precision@1 from 51.16\%$\pm$0.21 to 43.66\%$\pm$1.45. This indicates that vision-based structural refinements significantly enhance keyword alignment and retrieval accuracy.

In Dense Retrieval, the most significant drop is also observed with KAP w/o Vision, where Precision@1 declines from 53.65\%$\pm$2.24 to 42.98\%$\pm$0.65. Without vision-assisted layout parsing, the quality of semantic representations deteriorates, negatively impacting dense retrieval performance.

For Hybrid Retrieval, KAP w/o Vision again results in the largest performance loss, with Precision@1 decreasing from 59.73\%$\pm$1.10 to 47.33\%$\pm$0.81. The absence of vision-driven layout reconstruction undermines the integration of sparse and dense retrieval components.

These results collectively highlight that vision-based enhancements are critical across all retrieval settings. The consistent impact of vision removal confirms the importance of our multimodal approach, demonstrating that leveraging MLLMs for structural understanding substantially strengthens the overall retrieval effectiveness of KAP.

\section{Conclusion}

In this study, we introduced KAP, a two-stage preprocessing framework that improves text quality for Hybrid Retrieval in Traditional Chinese non-narrative documents. By integrating MLLM-based post-OCR processing, KAP effectively corrects OCR errors, restores table structures, and enhances text representation for both Sparse and Dense Retrieval.

Experimental results demonstrate that KAP significantly improves retrieval accuracy across all retrieval paradigms. The ablation study further validates the contributions of each component, confirming that OCR correction, layout-aware format reconstruction, and retrieval-aware rewriting collectively enhance retrieval performance.

Future work includes reducing the computational cost of the MLLM and extending KAP to other document domains. These improvements will further enhance the scalability and effectiveness of text preprocessing for retrieval tasks in real-world complex document settings.

\section*{Acknowledgments}
This study was supported by E.SUN Bank, which provided the dataset from the "AI CUP 2024 E.SUN Artificial Intelligence Open Competition." We sincerely appreciate E.SUN Bank for its generous data support, which has been invaluable to this research.

\bibliographystyle{ACM-Reference-Format}
\bibliography{ntcirsample}

\end{document}